\documentclass{entcs} \usepackage{prentcsmacro}

\usepackage{amsmath, amssymb, stmaryrd, wasysym, xspace}
\usepackage{graphicx}

\def\ens#1{\{#1\}}

\newcommand{\ket}[1]{\mbox{$|#1\rangle$}}
\newcommand{\gket}{\ket{\psi}}

\newcommand{\p}{\ket{+}}

\def\et#1#2{E_{#1#2}}

\def\nr{ \;|\; }
\def\given{\, \| \,}

\def\qci{\mathtt{qc?}} 
\def\cci{\mathtt{c?}} 
\def\qco{\mathtt{qc!}} 
\def\cco{\mathtt{c!}} 
\def\qcio{\mathtt{qc!?}} 
\def\ccio{\mathtt{c!?}}

\newcommand{\ox}{\otimes}

\def\qed{\ensuremath{\Box}}

\newcommand{\bA}{\ensuremath{\mathbf{A}}\xspace}
\newcommand{\bB}{\ensuremath{\mathbf{B}}\xspace}

\newcommand{\cE}{\ensuremath{\mathcal{E}}}

\newcommand{\cN}{\ensuremath{\mathcal{N}}}

\def\lastname{D'Hondt and Sadrzadeh}
\begin{document}

\begin{frontmatter}

  \title{Classical knowledge for quantum security ({\tt \large extended abstract})} 
     
\author{Ellie D'Hondt\thanksref{myemail}}
\address{Vrije Universiteit Brussel \& FWO, Belgium}

\author{Mehrnoosh Sadrzadeh \thanksref{vemail}}
\address{ Laboratoire Preuves Programmes et Syst\`{e}mes, Universit\'{e} Paris 7, France}

\thanks[vemail]{Email:
    \href{mailto:mehrs@comlab.ox.ac.uk} {\texttt{\normalshape
        mehrs@comlab.ox.ac.uk}}}
 \thanks[myemail]{Email:
    \href{mailto:Ellie.DHondt@vub.ac.be} {\texttt{\normalshape
        Ellie.DHondt@vub.ac.be}}}

\begin{abstract} 
We propose a decision procedure for analysing security of quantum cryptographic protocols, combining a classical algebraic rewrite system for knowledge with an operational semantics for quantum distributed computing. As  a test case, we use our procedure to reason about security properties of   a recently developed quantum secret sharing protocol that uses graph states. We  analyze three  different  scenarios based on the safety assumptions of the classical and quantum channels and  discover the path of an attack in the presence of an adversary. The epistemic analysis that leads to this  and similar types of attacks is purely  based on our classical notion of knowledge.
\end{abstract}


\end{frontmatter}

\section{Introduction}
Quantum communication is an inseparable  part  of  quantum computing: it offers solutions to the risks caused by the exponential speed-up in the power of  adversaries,  which is in turn caused  by quantum algorithms. While some advances have been made in the area of formal verification of quantum communication protocols~\cite{Gay05}, no applicable formal framework has yet been suggested for  their automatic cryptanalysis.  This is contrary to the fact that,  similar to the situation in classical security,  attacks have been  discovered on proven-to-be-safe quantum protocols.

In this paper, we present a  decision procedure  to verify wether a protocol satisfies an epistemic security property. Our procedure  derives  knowledge properties of agents   from  the set of  dynamic and epistemic traces of the protocol.  The \emph{dynamic traces}  are generated  from the protocol specification  by operational rules of   distributed measurement calculus (DMC)~\cite{Danos05b}.  These   are then expanded to the \emph{epistemic traces} using  \emph{appearances} of agents about the actions of the protocol. The appearances are derived from the  safety assumptions of the communication channels according to a set  of  rules.  Our  notions of knowledge and time are classical and have been used in  the formal analysis of classical  protocols, for example in the Halpern style models of~\cite{Prakash,Danos07} and in the algebraic Epistemic Systems of~\cite{BCS2,SadrThesis}. 

Both the DMC model and the algebra have been  previously used to  analyze the security of quantum key distribution (QKD) and its attacks~\cite{Danos07,DHondt05c,SadrAAAI}. 
The setting of this paper has advantages over both these attempts. First, we rely on the already existing rules of the semantics of DMC, as opposed to adding axioms for quantum mechanics to the algebra as pursued in~\cite{SadrAAAI}. Second, we use the algebraic axiomatics of dynamic and epistemic adjunctions  to derive knowledge properties of the protocol, as opposed to model-checking them by traversing the tree of the protocol as done in~\cite{Danos07,DHondt05c}. Third,  we set the actions of  the adversary in a compositional way using the appearance maps of the algebra, as opposed to ad-hocly adding them to the specification of the protocol as suggested in~\cite{Danos07,DHondt05c}. 

We prove that our decision procedure is sound and terminating with regard to the pair of a DMC model and the algebraic axiomatics of Epistemic Systems. We apply our decision procedure to   a new quantum secret sharing (QSS) protocol, which is  based on graph states and has been proposed recently in~\cite{Markham08}. For this protocol, we develop epistemic properties and prove them  for three kinds of  assumptions on the quantum channels: safe, unsafe with non-suspicious agents, and unsafe with suspicious agents. We show that in the second case, the protocol does not satisfy its desired epistemic property and is thus not secure, moreover, we discover the path of an intercept-exchange attack that caused this insecurity.  A full analysis of the safety assumptions  of all the channels and their impact on the security properties needs automation, which constitutes on going work. Also, we have only been working on a one-round basis and indeed, for a full analysis of protocols one needs to run the protocol in many runs and then use probabilities, for instance on the knowledge modalities. This would be a natural and exciting extension of the currently proposed framework.

In a nut shell, our framework  is obtained by merging the model checking approach of~\cite{DHondt05c,Danos07} and the algebraic axiomatics of~\cite{SadrAAAI}.   The former is based on a distributed extension~\cite{Danos05b} for an assembly language~\cite{Danos04b} that universally models computations of the one way model.  Its  knowledge operator is defined over Kripke  structures in the style of  Fagin et al~\cite{FH} by using equivalence relations on the states. Reasoning about properties of a protocol is done on the state space of this structure using a  logic with temporal and epistemic operators. 
The latter  is based on   the Stone-like duals  of these relational systems and moreover, following~\cite{CoeckeMS},  a quantale structure is assumed  on the actions. This setting consists of a pair of a quantale of classical and quantum actions and its right module of bits and qubits involved in a protocol. The pair is endowed with a family of join-preserving maps, one for each agent involved in the protocol.  The right adjoints  to these endomorphisms give rise to a very useful  notion of knowledge, both on propositions of module and actions of quantale. 

\section{Decision Procedure}
The procedure has three main steps. 
First, we write as program in the language of  the distributed measurement calculus (DMC) to implement the specification of the  quantum protocol  and generate a set of  \emph{dynamic traces} for it. This is done   by executing the rules of the operational semantics. of DMC   Second, we write  formulae with dynamic and epistemic modalities  to expresses security properties of the protocol. This is done  in the algebraic  syntax  of Epistemic Systems. Finally,  we apply an algebraic rewrite system to decide wether the protocol satisfies the properties. 

\smallskip
\noindent
{\bf Step (1)  Specify and produce dynamic traces in DMC.}\\
Programs of  DMC are implemented as \emph{networks of agents}. A {network of agents}  is denoted by $\cN$ and  is defined as follows 
\begin{center}\begin{minipage}{13cm}
\[
\cN= \gket \given \bA(Q).\cE\nr \bB(Q').\cE'\dots\quad\text{.}
\]\end{minipage}\end{center}

\medskip
\noindent
It consists of  a  set of agents acting in parallel (denoted by $|$)  on a given entanglement resource \gket.
An \emph{agent} $\bA(Q).\cE$ is specified by a name $\bA$, a set $Q$ of qubits it owns, and an event sequence $\cE$. The event sequence  can be a computation in the measurement calculus, a classical message reception $\cci x$ and sending $\cco y$, or  a qubit reception $\qci q$ and sending $\qco q'$. Note that, contrary to the original definitions in~\cite{Danos05b} we now write specifications from left to right; also agents may have extra classical parameters $a$, written as $A(a,Q)$.
As an example, here is one round of Ekert's implementation of QKD:
\begin{center}\begin{minipage}{13cm}
\[
QKD=\et 12 \| \bA(a,1).[H_{1}^{a};M_{1};\cco a; \cci b] \nr \bB(b,2).[H_{2}^{b};M_{2}; \cci a;\cco b] 
 \text{  .}
\]\end{minipage}\end{center}

\medskip
The set of traces of a program are generated by following the rules of the small-step semantics as specified in \cite{Danos05b}, but moreover, we work with projections, annotate actions with agents that performed them, and name the preparation actions of the initial entanglement resource \gket \ and  the distribution actions of qubits.  For example,  $P_i^{A,\alpha}$ stands for the spin $\alpha$ projection of qubit $i$ done by agent $A$ and $N_i^C$ is the preparation of qubit $i$ by agent $C$. The preparation actions are made explicit by juxtaposing them  to the left most of the traces; for QKD the
entanglement resource $\et 12$ is created by applying $N^C_1; N^C_2; E^C_{1 2}$ to a 2-qubit system $q_{1}\ox q_{2}$, where $N$ is preparation in the $\p$ state and $C$ is the agent who prepared the entanglement resource.  Distributing these qubits to agents $A$ and $B$ is denoted by a quantum broadcast action $\qcio^C_X q_i$, which stands for agent $C$ sending  qubit $q_i$ to agent $X$ and agent $X$ receiving it from him. This is a shorthand for a quantum send $\qco^C_X q_i$ and a quantum receive $\qci^X_C q_i$.  Similarly, we also shorthand a classical send $\cco^C_X  a$ and receive $\cci^X_C a$ to a broadcast $\ccio^C_X a$.

 According to these conventions two of the four possible traces for a successful run of QKD  become as follows

\begin{center}\begin{minipage}{13cm}
\[
\pi= N^C_1 ; N^C_2 ; E_{1,2}^{C} ;  \qcio^C_A\,  q_1; \qcio^C_B\,  q_2 ; P_1^{A, X} ; P_2^{B,X} ; \ccio^A_B\,  a ;  \ccio^B_A\,  b\quad\text{,}
\]

\medskip
\[
\pi'= N^C_1 ; N^C_2 ; E_{1,2}^{C} ;  \qcio^C_A\,  q_1; \qcio^C_B\,  q_2  ; P_1^{A, Z} ; P_2^{B,Z} ; \ccio^A_B\,  a ;  \ccio^B_A\,  b\quad \text{.}
\]\end{minipage}\end{center}

\bigskip
 \noindent
{\bf Step (2)  Write security properties in Epistemic Systems.}\\
The input to the rewrite  system is an expression of the form 
\[
l \vdash r
\]
 where $l$ is  the initial state and $r$ is  an epistemic property that  contains the disjunction of  dynamic traces  produced above. An example is the following expression  
\[
q_i \vdash [\pi]\Box_A \Box_A s_i^j
\]
which  is  read as
 
\begin{center}
 \fbox{\begin{minipage}{12cm}
 \begin{center} After running the trace $\pi$ of  the protocol on qubit $q_i$, agent $A$ knows that $B$ knows that the value of bit $i$ is $j$.\end{center}\end{minipage}}\end{center}
 
 \noindent
  The $l$ and $r$ expressions are   generated as follows:
\begin{center}\fbox{\begin{minipage}{13cm}
 \begin{itemize}
 \item The initial state  $l$ is made of  propositions $m$ that are formed by  closing  atomic classical and quantum variables $s_i^j$ and $q_i$ under $\neg, \wedge, \vee$ and logical constants $\bot, \top$. The variables are generated   as follows
 \[
 \kappa ::= s_i^j \mid q_l \mid q_l \otimes q_w
 \] 

\item The epistemic property $r$ is generated  as follows
\[
r ::= m \mid [\pi]m \mid \Box_A (m)
\]
where $\Box_A (m)$ is the epistemic modality and for $\pi$ a dynamic trace  $[\pi]m$ is the dynamic modality.
\end{itemize}\end{minipage}}\end{center}
One such expression for Ekert's QKD is
\begin{center}\begin{minipage}{13cm}
\[
q_1 \otimes q_2 \vdash \left[
 N^C_1 ; N^C_2 ; E_{1,2}^{C} ;  \qcio^C_A\,  q_1; \qcio^C_B\,  q_2 ; P_1^{A, X} ; P_2^{B,X} ; \ccio^A_B\,  a ;  \ccio^B_A\,  b
 \right]\Box_A \Box_B (s_1^0 \wedge s_2^0)
\]
\end{minipage}\end{center}

\medskip
\noindent
 Proving this property together with a permutation of it  for   $B$, that is
 \begin{center}\begin{minipage}{13cm}
\[
q_1 \otimes q_2 \vdash \left[
N^C_1 ; N^C_2 ; E_{1,2}^{C} ;  \qcio^C_A\,  q_1; \qcio^C_B\,  q_2 ; P_1^{A, X} ; P_2^{B,X} ; \ccio^A_B\,  a ;  \ccio^B_A\,  b
 \right]\Box_B \Box_A (s_1^0 \wedge s_2^0)
\]
\end{minipage}\end{center}
 will imply  that $A$ and $B$\emph{ share} a piece of data, which is the results of each other's measurements, that is $ (s_1^0 \wedge s_2^0)$. The sharing property is expressed by the nested knowledge property, that $A$ knows that $B$ knows it, and vice versa \footnote{It is arguable  wether one has to nest the knowledge modalities infinitely many times and thus use the common knowledge  operator to express the sharing property, but for now we restrict ourselves to a two level nesting.}. That the data is \emph{secret} is proved by showing that an adversary $E$ does not know it, that is the following expression
 \begin{center}\begin{minipage}{13cm}
\[
q_1 \otimes q_2 \vdash \left[
N^C_1 ; N^C_2 ; E_{1,2}^{C} ;  \qcio^C_A\,  q_1; \qcio^C_B\,  q_2 ; P_1^{A, X} ; P_2^{B,X} ; \ccio^A_B\,  a ;  \ccio^B_A\,  b
 \right] \neg \Box_E (s_1^0 \wedge s_2^0)
\]
\end{minipage}\end{center}

\bigskip
 \noindent
{\bf Step (3). Generate Epistemic traces and verify the property.}\\
We proceed by analyzing uncertainty of agents about the states and actions of protocols. These are referred to as  \emph{appearance} maps  and are denoted by $f_A$ for an agent $A$. They encode all possible actions or propositions that appear possible to an agent, given the action that is happeneing or the proposition that is true in reality, we refer the reader to~\cite{BCS2,SadrThesis} for discussions and examples. Here, we treat these maps  more practically and introduce a general set of rules to generate them. These  rules are presented  below.

\begin{center}
\begin{enumerate}
\item The  agents have no uncertainty about the steps of the protocol they are involved in.

\item Qubits are encoded as black boxes and thus appear as they are, that is as identity to all agents.  Classical bits appear as either 0 or 1 to agents.

\item  The owner of an action has no uncertainty about his actions, but is uncertain about other agents' actions. His appearances of these latter actions are generated by instantiating their  variables.

\item  There is only one adversary present in each protocol. This adversary can intercept the unsafe channels, either quantum or classical, by stopping the messages, changing the content of the messages, creating new messages and sending them. On a quantum channel, the change of the content of the message is done by   measuring the sent qubit and  the creation of new messages by preparing fresh qubits. On the classical channel, the change is simply affected by reading and writing the values of the bits.

\item On the safe channels, the adversary can either be passive or not present at all.  In the latter case, he cannot even see if messages are passing through and what is their content.  In the former case, on a classical channel, he can see the value of the bits  as well as the sender and receiver of each message, but cannot change anything. On a quantum channel, he can only see that a qubit is passing, but cannot see its state.

\item Communication actions on a safe channel are either public or private announcements to a subgroup of agents. The former appears as identity to all agents, whereas the latter is  identity only to the insiders in the group, and either as nothing or all possible choices to the outsider agents. On an unsafe channel, the announcement actions are treated as   separate send and receive actions.

\item Honest agents may suspect the interception actions of the adversary. If they do so, these actions  appear to them as either happened or not. If they do not, they appear to them as the neutral action in which nothing happens.
\end{enumerate}\end{center}

\medskip
\noindent
For example, the appearances of the projection action $P_1^{A, X}$ in our above example traces are  as follows

\begin{minipage}{13cm}
\[
f_A(P_1^{A, X}) = P_1^{A, X}, \qquad f_B(P_1^{A, X}) = P_1^{A, X} \vee P_1^{A, -X} \vee P_1^{A, Z} \vee P_1^{A, -Z}\quad .
\]
\end{minipage}

\medskip
\noindent
The appearances of the communication actions depend on the safety assumptions of the channel in which they take place. For example, if the channel is safe, they are treated as broadcasts otherwise as separate send and receive actions.  We present  a detailed example on those in the last section. 

\medskip
\noindent
Due to space limits we cannot present the rewrite rules;  they are similar to the system  presented in~\cite{SimonSadr}. By applying them, one first  eliminates the    logical connectives  $\wedge, \vee,\Box_A, [\  ]$ and  then the classical  and quantum communication actions. The output is a  set of \emph{atomic expressions}, defined as follows
\begin{definition}
An expression  $l  \vdash r$ is atomic iff $l$ is a   quantum state followed by a sequence of atomic quantum actions and $r$ is an atomic classical or quantum state. 
\end{definition}

\noindent
For instance, for a safe quantum channel, the atomic form of the our sharing property is 
\[
(q_1 \otimes q_2)  (N^C_1 ; N^C_2 ; E_{1,2}^{C} ; P_1^{A, X} ; P_2^{B,X}) \vdash s_1^0 \wedge s_2^0
\]
These  atomic expressions may contain new epistemic uncertainties and thus will need to be  verified against  our  operational semantics.  For this purpose, we introduce below  the notion of a \emph{well-defined} expression.
\begin{definition}
An atomic expression $l \vdash r $ is \emph{well-defined} iff $l$ is derivable within the operational semantics of DMC. It is true iff $r$ holds in all configurations resulting from $l$. An epistemic property holds for a protocol whenever  all its well-defined atomic expressions are true. 
\end{definition}

\begin{proposition}
For a protocol specification $\cN$ and an expression $l \vdash r$ which is built  from the dynamic traces of $\cN$, the process of deciding if the epistemic property in $r$ holds for $\cN$ is terminating  and  sound  with regard to the pair of an Epistemic System  and a DMC model.  
\end{proposition}
\noindent
{\bf  Proof.} These  follow from image finiteness of appearances of actions and propositions, together with  soundness and termination of the rewrite system of Epistemic Systems and the DMC model~\cite{Danos05b,SimonSadr}.  

 
 \section{Case study: quantum secret sharing}

We apply our procedure  to the quantum secret sharing (QSS) protocol recently established in \cite{Markham08}. In secret sharing a dealer holds a secret bit which he wants to send to $n$ players, such that at least $k$ players are needed to reconstruct the secret. The problem is well-known in the  classical settings and solvable for all $(n,k)$. In the quantum case, only the  $(n, n)$ case has been solved for the  GHZ-type entanglement \cite{Xiao04}. The work in \cite{Markham08} uses instead graph states and thus is more suitable for modelling  in our measurement-based setting. Moreover, it generalizes  the quantum key distribution protocols and simplifies their proofs. We analyze and prove some of the epistemic properties of the QKS component of the $(3,5)$ case, where a particular graph state is used to establish a secret key between three players and the dealer in one go (as opposed to via several 2-party QKD protocols). This key will then be used to distribute a secret using the other components of the protocol. 

\begin{center}
\scalebox{.60}{\includegraphics{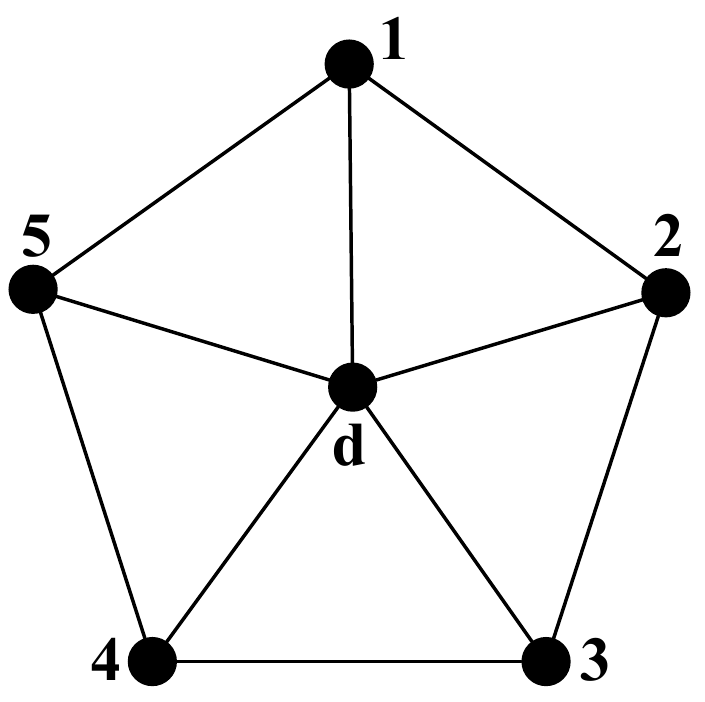}}
\end{center}
The recource required for the protocol is the graph state shown above, henceforward called $G(3,5)$. It is prepared following the usual procedure for graph states, that is
\begin{center}
\begin{minipage}{13cm}
 \[
G(3,5)= (N_{1};\dots ;N_{5};N_{6};\prod_{e_{ij}}E_{ij})  \otimes_{i= 1}^6 q_i \\; \text{,}
 \]
 \end{minipage}\end{center}

 \noindent
where  $e_{ij}$ is the set of edges. The protocol proceeds as follows:

\hspace{-0.8cm}\fbox{\begin{minipage}{14cm}
\begin{description}
\item[Step 1.] The dealer prepares $G(3,5)$, sends each agent a qubit $q_{i}$ together with an agent identity $i$.

\item [Step 2.] The dealer measures his qubit in the $Y$ or $Z$ basis randomly and broadcasts his measurement basis.

\item [Step 3.] Each participating player measures his qubit  in the $X$,  $Y$ or $Z$ basis randomly, then broadcasts his identity and measurement basis.

\item [Step 4.] Depending on these messages, each agent determines if the run was successful as follows: 
\begin{itemize}
\item If the participating agents are neighbours, then we have $ijk=i(i+1)(i+2)$; this is the case for  the following  measurement combinations 
\[
M_{6}^{Z}M_{i}^{Z}M_{j}^{X}M_{k}^{Z} \quad \text{and} \quad M_{6}^{Y}M_{i}^{X}M_{j}^{Y}M_{k}^{X} \; \text{.}
\]
\item If they are in a so-called T-shape,  we have $ijk=i(i+1)(i+3)$; this is the case for the following measurement combinations 
\[
M_{6}^{Z}M_{i}^{X}M_{j}^{Y}M_{k}^{Y}  \quad \text{and} \quad M_{6}^{Y}M_{i}^{Y}M_{j}^{Z}M_{k}^{Z} \; \text{.}
\]
\end{itemize}
\item [Step 5.] For a successful run, measurement outcomes are correlated as $
s_{6}=s_{i}\oplus s_{j}\oplus s_{k}$.
Players use their secure classical channels to exchange measurement outcomes and determine if $s=s_{6}$, hence establishing a shared key with the dealer. 
\end{description}\end{minipage}}

\smallskip
We refrain from giving the full specification of the QSS network and move straight on to its traces, where we  treat all the communication actions as broadcasts and later on  break them to separate send and receive actions, as necessary and according to the safety assumptions of their channels.   Whenever the subscript of a broadcast action is missing, e.g. in $\ccio^D a$ it means that the broadcast is a public action that can be listened to by everyone.  A typical trace for a successful run of QSS is as follows

\begin{center}
\begin{minipage}{13cm}
$$
\begin{array}{rll}
\pi = & N^D_{1};\dots ;N^D_{6};\prod_{e_{ij}}E^D_{ij}&\text{(preparation)} \\
&(\qcio^D_1  q_{1})\dots(\qcio^D_5 q_{5}) &\text{(private broadcast of qubits)}\\
& P_6^{D, \pm a}P_i^{ A_i, \pm b}P_j^{ A_j, \pm c}P_k^{A_k,\pm f}&\text{(measurement projections)}\\
& \ccio^D a ; \ccio^{A_i} b ; \ccio^{A_j} c ; \ccio^{A_k} c &\text{(public broadcast of measurement bases)}\\
& (\ccio^{A_i}_{A_j,A_k} s_{i})(\ccio^{A_j}_{A_i,A_k} s_{j})(\ccio^{A_k}_{A_i,A_j} s_{k})&\text{(private broadcast of player's mes. outcomes)}.
\end{array}
$$
\end{minipage}\end{center}

\medskip
\noindent
Here $a \in \ens{X,Y}, b,c,f \in \ens{X,Y,Z}$ are measurement basis, $\qcio^{D}_{i} $ is the quantum message passing from $D$ to $A_i \in \{A_1, \cdots, A_5\}$ denoting the 5 players, and $\ccio^{A_i}_{\beta}$ is the private announcement from player $A_i$ to the group $\beta \subseteq \{A_1, \cdots, A_5\}$. We omit  the calculation of the secret key, which is determined  by the following exclusive-or  formula 
\[
s=s_{i}\oplus s_{j}\oplus s_{k}
\]
 Successful traces only depend on the chosen values for $a,b,c$ and $f$; one example of such a trace for adjoining agents $A_1$, $A_2$ and $A_3$, owning qubits 1,2 and 3 respectively, is  as follows

\begin{minipage}{13cm}
\begin{align*}
\pi = \dots&\\
& P_6^{D,+Z}P_1^{A_1,-Z}P_2^{A_2,-X}P_3^{A_3,+Z};\\
& Z!^{D} ; Z!^{A_1} ; X!^{A_2} ; Z!^{A_3};\\
&(\ccio^{A_1}_{A_2,A_3} \; 1)(\ccio^{A_2}_{A_1,A_3} \; 1)(\ccio^{A_3} _{A_1.A_2} \; 0)&
\end{align*}
\end{minipage}

\subsection{Epistemic Properties}
We consider three cases: agents' heaven,   adversary's heaven, and adversary's hell. In the first case the quantum channel is safe, in the second case it is not and the honest agents do not suspect it, in the third case it is not and the honest agents do suspect it. The other channels are assumed to be safe in all three cases.  For each case, we show  how the appearances of agents of actions in the dynamic traces are set. This  is  done according to the safety assumptions on the channel and our rules.  Then we present some of the related epistemic security properties  of each case. 
\begin{enumerate}
\item {\bf Agents' heaven}\\
The appearance of the projections are set according to the rule $(iii)$ of appearances.  Since the channels are safe,  the communication actions on the quantum channel are  treated as public broadcasts and  by rule $(vi)$ and for $\sigma$ an agent they are set as follows
\[
f_{\sigma}(\qcio^{D} q_i) =   \qcio^{D} q_i
\]
That is, all the agents are fully aware of the broadcast action and thus have only one possibility in their appearance, the broadcast action itself. 
 The communication actions on the classical channels are  private announcements and by rule $(vi)$ their appearances are set as follows,  for  $\beta$ a  subset of players 
\begin{center}
\begin{minipage}{13cm}
\[
f_{\sigma} (\ccio _{\beta}^{A_i} s_i^j) = 
\begin{cases}
\ccio _{\beta}^{A_i} s_i^j & \sigma \in \beta\\
&\\
\ccio_{\beta}^{A_i} s_i^j \vee \ccio _{\beta}^{A_i} \overline{s}_i^j & \sigma \notin \beta
\end{cases}
\]\end{minipage}\end{center}
This says that the insider agents $\sigma \in \beta$ who receive the sent  bit $s_i^j$, which is either equal to 1 or 0, are fully aware what has happened and thus have only one possibility about the private broadcast action,  that is  the broadcast action itself and thus their appearance is identity. But by rule $(i)$ of appearances, the outsider agents $\sigma \notin \beta$ are only aware that a bit has been privately broadcasted to the subgroup $\beta$ and are uncertain about the value of that bit.  So they consider it possible  that either a bit with value 1 or a bit with value 0 has been privately broadcasted to insider agents in $\beta$. Thus  their appearance is the choice of these two possibilities.

Some of the  epistemic properties of interest for our trace $\pi$,  allied players $A_i \in \{A_1,A_2, A_3\}$,  joined with dealer $ \sigma \in \{D, A_1,A_2, A_3\}$   are as follows

\medskip
\hspace{-0.3cm}\fbox{
\begin{minipage}{12.2cm}
\begin{itemize}
\item The dealer knows his bit and  binary sum of allied players' bits, i.e.
\[
\Box_D\,  (s_6^0 \wedge (s_1^{b_1}\oplus  s_2^{b_2} \oplus s_3^{b_3}))\,.
\]

\item Allied players moreover know the value of each single measurement,  i.e.
\[
\Box_{A_i}\, ( s_6^0 \wedge s_1^{1} \wedge s_2^{1} \wedge s_3^{0})\,.
\]

\item The dealer knows  that the players know his bit and the players know that the dealer knows the sum of their bits,  i.e.
\[
\Box_D \Box_{A_i}\,  s_6^0 \quad \text{and}\quad \Box_{A_i} \Box_D   (s_1^{b_1}\oplus  s_2^{b_2} \oplus s_3^{b_3})\,.
\]

\item The adversary does not know any of the above,  i.e.
\[
\neg \Box_E\,  (s_6^0 \wedge (s_1^{b_1}\oplus  s_2^{b_2} \oplus s_3^{b_3}))\,.
\]

\item The dealer and the agents know the above,  i.e.
\[
\Box_{\sigma} \neg \Box_E\,  (s_6^0 \wedge (s_1^{b_1}\oplus  s_2^{b_2} \oplus s_3^{b_3}))\,.
\]
\end{itemize}
\end{minipage}
}

\vspace{5mm}
\item {\bf Adversary's heaven}\\
In this case, the quantum channel is not safe and  by rule $(iv)$ the adversary  can intercept the channel.  By rule $(vi)$ since the channel is not safe, we  must break  its broadcasts to  separate send and receive actions. The  appearances of these actions to the agents involved in them (e.g. the appearance of the sent action to the agents  who received it) are not identities any more.   The appearances for  the send of a qubit   are set as follows, where $q_j$ is a new qubit with  $j \geq 7$
\begin{center}
\begin{minipage}{13cm}
\[
f_{\sigma'} (\qco ^D_i q_i) =  
\begin{cases}
\qco ^D_i\,  q_i ; \qci ^E_D\,  q_i ; P_i^{E,e};  N_j^{E,e} ; \qco ^E_i\,  q_j & \sigma' = E\\
&\\
\qco ^D_i q_i & o.w.
\end{cases}
\]\end{minipage}\end{center}
This says that neither the agents nor the dealer  suspect that $E$ intercepted dealer's sent qubit and thus their appearance of dealer's sent action is (wrongly) identity. Where as in reality, $E$ did the following sequence of interception events, but they only appear to him as identity.
\[
\qco ^D_i\,  q_i ; \qci ^E_D\,  q_i ; P_i^{E,e};  N_j^{E,e} ; \qco ^E_i q_j  
\]
According to this sequence of events, $E$ received the dealer's sent qubit $q_i$ that was meant to be received by agent $i$, measured it, then  prepared a corresponding new qubit $q_j$ and sent it to agent $i$.  For the corresponding receive action, it appears to the dealer  that players received the qubit that he sent to them, $f_D (\qci ^{A_i}_D\,  q_i)=\qci ^{A_i}_D\,  q_i$, whereas in reality they receive the qubit sent to them by adversary, $f_{A_i} (\qci ^{A_i}_D\,  q_i)=\qci ^{A_i}_E\,  q_j$.
In case the eavesdropper is lucky and chooses the right projection for all three qubits he intercepts, he is able to derive the value of the key. In this case some of the  epistemic properties of interest are

\medskip
\hspace{-0.6cm}\fbox{
\begin{minipage}{13cm}
\begin{itemize}
\item   The adversary knows the shared key, i.e. $\Box_E s_6^0 $.

\item The  players and the dealer wrongly think that he does not know  this, i.e. 
\[
\Box_{\sigma}  \neg \Box_E s_6^0\,.
\]
\end{itemize}
\end{minipage}}

\medskip
\noindent
Note that here the adversary has to be more  lucky than in Ekert'91. This is because he has to intercept the qubits of three allied  players instead of one, and has to choose from three measurement bases.

\vspace{5mm}
\item {\bf Adversary's hell}\\
This is the same as above, but the players suspect adversary's actions, that is according to rule $(vii)$,  it appears to them either there was no interception or  there was one and the above sequence of actions took place by the adversary. Thus we obtain

\medskip
\begin{minipage}{13cm}
\[
f_{A_i} (\qco ^D_i\,  q_i) =  \qco ^D_i\,  q_i  \vee (\qco ^D_i\,  q_i ; \qci ^E_D\,  q_i ; P_i^{E,e} ;  N_j^{E,e} ; \qco ^E_i\,  q_j )
\]
\end{minipage}

\noindent
\medskip
Similarly, the dealer suspect adversary's actions on the receipt of his sent qubit
\begin{center}
\begin{minipage}{13cm}
\[f_D(\qci ^{A_i}_D\, q_i) = \qci ^{A_i}_D\, q_i \vee (\qci ^E_D\, q_i ; P_i^{E,e};  N_j^{E,e} ; \qco ^E_i q_j ; \qci ^{A_i}_E q_j)
\]\end{minipage}\end{center}

\medskip
\noindent
In this case, an interesting epistemic property would be the following 

\medskip
\begin{center}\fbox{
\begin{minipage}{12cm}
\begin{center} The dealer and the players are  not sure anymore if the adversary knows their secret bit, and thus if the bit can be treated as a secret i.e. \end{center}
\[
\neg \Box_{\sigma} \neg \Box_E s_6^0\,.
\]
\end{minipage}}\end{center}
\end{enumerate}

\subsection{Verifying Epistemic Properties}
As  examples, we verify two properties: one from the agents' heaven  and one from the adversary's hell. 

\begin{itemize}
\item {\bf Agents' heaven} \\From this scenario, we verify the following property
\[
\otimes_{i= 1}^6 q_i \vdash [\pi] \Box_D \Box_i\,  s_6^0
\]
The atomic expressions are generated via the following rewritings, where $\alpha_i$'s denote the juxtaposed actions of $\pi$

\begin{center}
\hspace{-1.4cm}{\begin{minipage}{15cm}
\begin{align*}
\otimes_{i= 1}^6 q_i \vdash [\pi] \Box_D \Box_{A_i}\,  s_6^0& \quad  \leadsto\quad 
\otimes_{i= 1}^6 q_i ; \pi \vdash\Box_D \Box_{A_i}\,  s_6^0 \quad \leadsto \quad \\
f_{A_i} f_D(\otimes_{i= 1}^6 q_i ; \pi ) \vdash s_6^0  & \quad \leadsto \quad
 f_{A_i} f_D(\otimes_{i= 1}^6 q_i) ; f_{A_i} f_{D}(\pi)  \vdash s_6^0 \\
 &\quad \leadsto\quad 
 f_{A_i} f_{D}(\otimes_{i= 1}^6 q_i) ; f_{A_i} f_{D}(\alpha_1); \cdots ; f_{A_i} f_D(\alpha_n)  \vdash s_6^0.
\end{align*}
\end{minipage}}\end{center}

\medskip
\noindent
By rule $(ii)$ of appearances we have $f_{A_i} f_{D}(\otimes_{i= 1}^6 q_i)  = \otimes_{i= 1}^6 q_i$. By rule $(iv)$ and our assumptions on channels,  we have $f_D(\alpha_i) = \alpha_i$ for $\alpha_{A_i}$ a quantum or broadcast communication action. By rule $(vi)$ for  communication between players we have $f_D(\ccio _{\beta}^{A_i} s_i^j) = \ccio _{\beta}^{A_i} s_i^j \vee \ccio _{\beta}^{A_i} \overline{s}_{i}^j$. Similarly for the projection actions we have

\begin{center}
\begin{minipage}{13cm}
\[f_D(P_6^{D,+Z}) = P_6^{D,+Z}
\]
and
\[
 f_D(P_1^{A_1,-Z})  = P_1^{A_1,-Z} \vee P_1^{A_1,+Z} \vee P_1^{A_1,-X} \vee P_1^{A_1,+X} \vee P_1^{A_1,-Y} \vee P_1^{A_1,+Y}\quad \text{.}
\]
\end{minipage}\end{center}

\medskip
\noindent
The values for the $f_{A_i}$'s are similarly set.   Substituting these values in the above expression, we first eliminate the traces in which the bases of projections do not match the announced  bases. Next we eliminate the communication actions from these traces whose content do not match the projections. As a result, we obtain a set of atomic expressions, of which only those satisfying $s_{6}^{0}=s_1^{b_1}\oplus  s_2^{b_2} \oplus s_3^{b_3}$ are well-defined in DMC. An example (out of  four) is
\begin{center}
\begin{minipage}{13cm}
\[
\otimes_{i= 1}^6 q_i ; N^D_{1};\dots ;N^D_{6};\prod_{e_{ij}}E^D_{ij} ; P_6^{D,+Z}; P_1^{A_1,+Z}; P_2^{A_2,-X}; P_3^{A_3,-Z}  \vdash  s_6^0\quad \text{.}
\]
\end{minipage}\end{center}
This atomic expression is true, since in all its final configurations  $s_6$  is 0, and thus our epistemic property holds for the secret sharing protocol.  

\item
{\bf Adversary's hell}\\
On the contrary, in the adversary's hell, one    shows that the epistemic property  $\Box_D \neg \Box_E s_6^0$ does not hold and thus $s_6^0$ is not treated a secret anymore. Moreover, we also discover paths of an intercept-change attack for each agent, for example the one for the player $A_1$ contains the following sequence of actions
\begin{center}
\begin{minipage}{13cm}
\[\cdots ; \qco ^D_1\, q_1 ; \qci ^E_1\, q_1; P_1^{E,+Z};  N_7^{E,+Z} ; \qco ^E_1\, q_7; \qci ^{A_1}_E\, q_7; P_7^{A_1,+Z}; \cdots
\]\end{minipage}\end{center}
In this path, the adversary receives dealer's original qubit $q_1$ that was meant to be received by agent 1, then measures it in basis $Z$ by doing projection $P_1^{E,+Z}$, then prepares a new qubit $q_7$ according to his measurement result and sends it to agent 1. The adversary turns out to be lucky and agent 1 picks the same measurement basis as him, that is $Z$ and does the same projections. The classical result of this projection   will for sure be the same as adversary's but might not the same as the dealer's.   
\end{itemize}

 \section{Conclusion}
In this article we proposed a new framework for formal analysis of security issues in quantum cryptographic protocols. Our  framework combines an algebraic rewrite system with a specification language for quantum distributed computations. The former provides machinery to work with uncertainties of agents in a protocol in a compositional way, while the latter inherently encodes the rules of quantum mechanics. Our framework was put to test in the analysis of a recent quantum secret sharing protocol based on graph states, where we proved some epistemic properties of the protocol in the presence and absence of an active adversary and discovered paths of an intercept-exchange attack. 
 For a full analysis one needs to generate many more epistemic traces and the need for automation and software implementation  is gravely felt.   A software implementation of the algebra~\cite{SimonSadr} is already in place to handle part of the verification. The construction of a tool that automatically derives the traces and semantics of a protocol is currently underway.

\section*{Acknowledgments}
 We had fruitful discussions with V. Danos, P. Panangaden,  D. Markham.  The second author has given talks  on a version of the algebra with some quantum axioms encoded in it, which was  partly based on joint work with E. Kashefi.

\bibliographystyle{entcs}
\bibliography{references}

\end{document}